# Geometrical characterization of healthy red blood cells using digital holographic microscopy and parametric shape models for biophysical studies and diagnostic applications.

Gaurav D. Bhabhor[1], Chetna Patel[2], Nishant Chhillar[3], Arun Anand[4,*], Kirit N. Lad[4,*]

[1] Department of Physics, Arts, Science & R.A. Patel Commerce College, Bhadran-388 530, Anand, Gujarat, INDIA

[2] Department of Physics, V. P. and R. P. T. P. Science College, Vallabh Vidyanagar-388120, Gujarat, INDIA

[3] Department of Physics, National Institute of Technology Delhi, New Delhi, INDIA.

[4] Department of Physics, Sardar Patel University, Vallabh Vidyanagar-388120, Gujarat, INDIA

## Abstract

Modeling of the red blood cell (RBC) shape is an integral part of the experimental and computer simulation investigations of light scattering by these cells for fundamental studies as well as diagnostic applications in techniques like cytometry and quantitative phase imaging. In the present work, a comprehensive study of the geometrical characterization of healthy human RBCs using digital holographic microscopy (DHM) and six frequently employed parametric shape models is reported. It is shown that the comparison of the optical phase profiles, and the thickness profiles given by the models with the DHM results gives a better judgment of the appropriateness of the parametric shape models. It is also shown that the RBC parametric models offer a simpler solution to RI-thickness decoupling problem in QPI methods. Results of geometrical characterization of 500 healthy RBCs in terms of volume, surface area, and sphericity index led to the classification of the parametric models in two categories based on the nature of variation of these quantities with the cell diameter. In light of the variability of the healthy RBC shapes, our findings suggest that the parametric models exhibiting a negative correlation between the sphericity index and the cell diameter would provide more reliable estimates of the RBC parameters in diagnostic applications. Statistical distributions and descriptive statistics of the RBC volume, surface area and sphericity index serve as a guide for the assessment of the capability of the studied parametric models to give a reliable account of the variability of the healthy RBC shape and size.

---

[*] Corresponding authors: knlad-phy@spuvvn.edu ; aanand-phy@spuvvn.edu

# 1. INTRODUCTION

The biconcave discoid shape of the human RBC is central to its main physiological function of the to-and-from transport of $O_2$ and $CO_2$ gases from the lungs to the tissue cells. The biconcave discoid shape is essential for the flexibility of the cell membrane to undergo large deformations without losing its biomechanical properties during its transit through the narrow capillaries. Abnormalities in the cytoskeleton and the cytoplasm of the RBCs, due to diseases (like malaria) or disorders (such as sickle cell anemia), impair its deformability and cause an irreversible change in the shape in the equilibrium fluidic environment of the blood. Thus, the knowledge of the geometrical shape and size parameters of RBCs plays an important role in their biophysical characterization and the diagnosis of related diseases and disorders. Cytometry techniques such as scanning flow cytometry [1-4] and a variety of quantitative phase imaging (QPI) methods using digital holographic microscopy (DHM) [5-15] have been developed in the last two decades for the investigation of the shape and geometry of RBCs for differentiating the subpopulations of the cells for diagnosis of disease and disorder. The latter techniques are often interchangeably termed as quantitative phase microscopy (QPM) [5-7] and, include common-path diffraction optical tomography[16], and defocusing microscopy [17]. Combinatorial methods like diffraction phase cytometry [18], tomographic flow cytometry[19], and polarization diffraction imaging flow cytometry [20] have also been developed. The diagnostic accuracy of these techniques relies on their ability to distinguish between the populations of healthy RBCs and those with impaired functionalities on account of irreversible shape alterations due to a disease or a disorder. The task of discriminating RBC populations of different shapes requires an accurate correlation between the shape and the geometrical parameters of the RBCs. For example, the correlation between the thickness distribution in the healthy RBCs and the malaria-infected RBCs has been used for their automatic classification using digital holographic interferometric microscopy.[21] It has also been demonstrated that joint statistical distribution of the characteristic parameters (say-surface area (SA), volume, sphericity index (SI), etc.) of RBCs, obtained from holographic QPI, can be used as feature patterns to classify RBC populations with different shapes and Hb content.[9] Shape parametrization has also been shown to be important for the biomechanical characterization of the cell membrane using 3D confocal microscopy.[22]

Computer simulation and parametrization of the cell shapes are often an integral part of many of the experimental characterization and diagnostic techniques. For example, in flow cytometry, computer simulation of light scattering by a single RBC employs a cell shape generated using different numerical methods [23-25] or parametric analytical models [1, 26-37]. As the distribution of forward light scattering depends on the size, shape, and hemoglobin (Hb) content of RBCs, the extraction of accurate information of these parameters from the light scattering profile (LSP) in the cytometry has been turned into an inverse problem of fitting the obtained LSP to those generated using computer simulations.[3, 38-40] The RBC volume distribution obtained through this approach gives the red cell distribution width (RDW), one of the complete blood count clinical test indices.[41] Iterative numerical methods of construction of RBC shape, based on membrane bending energy minimization, are not only computationally intensive but, also lead to the ambiguity related to two or more different shapes corresponding to the minimum membrane bending energy for the same set of characteristic geometrical parameters such as the volume (V) and the sphericity index (SI). To avoid such ambiguities, parametric models are most often used for light scattering simulation from single RBCs.[3] RBC shape parametrization is also important in many fundamental simulation studies related to cell membrane properties. A study to test the hypothesis that the shear elasticity arising from the membrane-associated cytoskeleton is necessary to account for shapes of real RBCs compares the theoretical cell surfaces generated using spherical harmonic series expansions and those estimated from 3D confocal microscopy images of live cells.[22] A surface model of RBC has been used to simulate changes in membrane curvature under strain.[36] A multiscale RBC model proposed to describe accurate mechanics, rheology, and dynamics uses a parametric model to generate an initial equilibrium RBC shape.[42] Parametric RBC shape models are also used in numerical simulations of dielectric spectra of RBC suspensions for the study of passive electrical properties of the cell membrane.[43-48] A study of the effect of external electromagnetic field on the shape of the RBCs demonstrates the utility of a parametric equation in the calculation of the induced transmembrane potential.[37] Another noteworthy utility of the parametric models is found in the QPM for testing the convergence and robustness of the algorithms [49] and; the applicability of the methods for retrieval of the refractive index (RI) of the cells from the phase images[50].

Considering the prevalent use of the parametric RBC models in numerical simulation studies as well as in diagnostic techniques, the present study focuses on two objectives: (1) Experimental characterization of the geometrical shape and size of healthy RBCs through QPI using DHM so that a thorough investigation of the parametric RBC shape models can be carried out from the viewpoint of their ability to generate the correct equilibrium shapes with the variability of the key geometrical parameters (volume, SA, and SI) within the experimentally reported normal range and, (2) Explore the utility of the parametric models to extract the RI profile from the phase profile obtained through QPI. RI profiles of RBCs are important for deciphering the information about the heterogeneities in the cell cytoplasm. Owing to the coupling of RI and the thickness in the QPI methods, the RI profiles are usually obtained using sophisticated optical tomography techniques [51-54] or digital holographic techniques that use algorithms to process multiple optical phase maps and algorithms [49,55-56]. Thus, *a priori* knowledge of the 3D shape of the cell from a parametric model could provide a simpler solution to the RI-thickness decoupling problem [49,54] in the QPI methods. We also look into the fundamental aspects like how the different parametrization in the RBC shape models results in qualitatively different functional dependence of the volume, SA, and SI on the diameter of the RBCs. From the several parametric models [1,26-37] that have been proposed over the years, we choose six models that include the classic Cassini model [26,27], the models proposed by Fung et al [28-30], Skalak et al [31,32], Kuchell and Fackerell [35,36], Yurkin et al [1] and San Martin et al [37]. Our choice of these models is based on how often the models are being used in various numerical simulation studies, and their general applicability for modeling different RBC shapes such as stomatocyte, and echinocyte. To emphasize the significance of the detailed investigation of RBC shape models undertaken in the present work, we would like to mention that the only comprehensive study of different RBC shape models by Valchev et al [57] is restricted to the comparison of meridional contours of the equilibrium RBC shapes to the two-dimensional (2D) RBC profile traces from normal optical microscope photographs reported by Jay [58] in 1975.

## 2. MATERIALS AND METHODS

### 2.1 Digital Holographic Microscopy and Quantitative Phase Imaging

*2.1.1 Sample preparation*

Blood samples were collected from a 26-year-old male donor with a B+ blood group using the pinprick method. The blood drop was directly mixed with 10 mL of 0.9% normal saline solution to ensure uniform tonicity and refractive index. A dropper transferred a drop of this mixture onto a clean glass slide, which was then covered with a cover glass. The slides were examined with a digital holographic microscope at room temperature within an hour of the sample collection.

*2.1.2 Experimental setup*

A schematic set up of a digital holographic microscope used for 3D imaging of RBCs is shown in Fig. 1(a).[21, 59-60] A low-power He-Ne laser with wavelength 632.8 nm and maximum output power of 2 mW is used as a light source. The input laser beam is split into two using a 50:50 beamsplitter. The transmitted beam from the beamsplitter is made to transilluminate the sample using a front-coated mirror and act as the object beam. The sample is kept on a translation stage for axial positioning. Nikon Achromatic 40x (NA=0.65) objective lens (MO1), then creates a magnified image of the sample on the hologram recording sensor (Thorlabs CCD array, 8-bit monochrome, 1024 x 768 pixels, 4.65 μm pixel pitch). The portion of the incident wavefront reflected by the first beam splitter acts as the reference beam, making a vertical microscope setup ideal for imaging cells immersed in fluids. The refence beam also passes through an objective lens (MO2) with a similar configuration as the one used to magnify the sample. The CCD array is located at the image plane of MO1, so that the hologram plane and image plane coincide (image plane digital holographic microscopy) [21, 59-60]. Since MO2 is also located exactly at the same distance as MO1 from the CCD array, the refence wavefront is an expanding spherical wavefront of the same curvature as that of object wavefront. This matching of curvatures of the object and reference wavefronts results in the generation of holograms containing linear interference pattern (Fig. 1c and 1e), which is helpful during the numerical reconstruction process. Since the present case involves image plane holography, the numerical reconstruction process only involves Fourier fringe analysis [61] rather than propagation of the diffracted wavefront [21, 59-60].

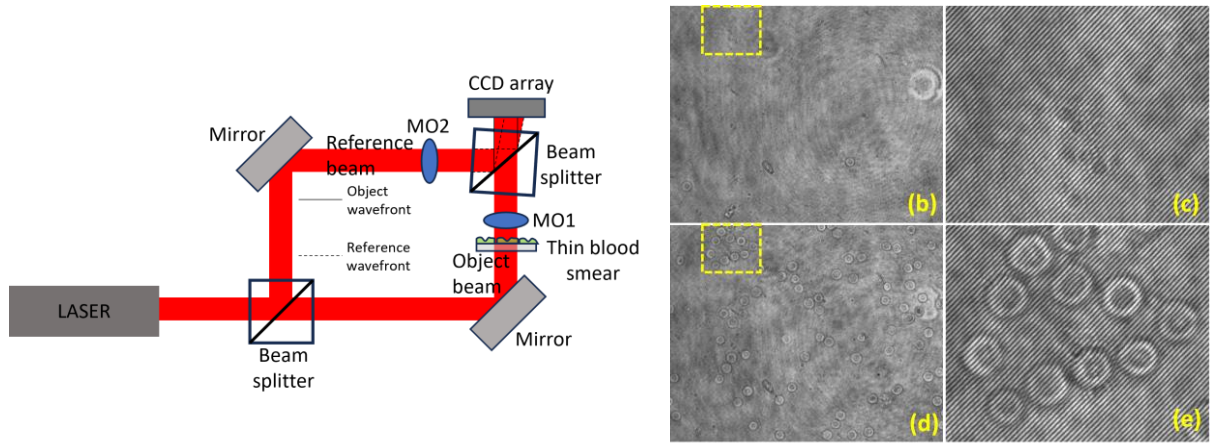

**FIGURE 1:** (a) Schematic diagram of the Digital Holographic Interferometry Microscope. (b) Reference hologram without object (RBCs) (c) An enlarged view of the selected segment of the reference hologram, (d) Hologram with objects (normal RBCs) and the background (0.9% saline solution), (e) An enlarged view of the selected segment of the object hologram.

### 2.1.3 Image Reconstruction

3D profiling of the RBCs requires extraction of the phase information from the recorded digital holograms. To extract the phase information, two holograms are recorded, one with the red blood cells and the surrounding medium (0.9% saline solution) in the field of view, called an object hologram $H_O$ (Fig. 1d and 1e) and one recorded with only the medium (saline solution) surrounding the red blood cells in the field of view called the reference hologram $H_R$ (Fig. 1b and 1c). Numerical reconstruction of digital holograms works by simulating the propagation of the digitally inputted reference beam from the recorded interference fringes [62] using angular spectrum propagation (ASP) diffraction integral [63], which is suitable for short-distance propagations. The numerical reconstruction of hologram using ASP diffraction integral can be mathematically represented as [64]

$$U(x,y,d) = F^{-1}\left\{filt[F\{H(x_h,y_h,0)R(x_h,y_h,0)\}]\,e^{ikd\sqrt{1-\lambda^2 f_X^2 - \lambda^2 f_Y^2}}\right\} \qquad (1)$$

where $U(x, y, z=d)$ is the complex amplitude distribution of the magnified object at the best focus plane situated at a distance $d$ from the hologram plane at $(x_h, y_h, z=0)$. On the right-hand side of Eq. (1), $F$ represents the Fourier transform of the hologram $h$ multiplied by the digital version of the reference wavefront. The exponential term represents the Fourier transform of the free space propagation function where $f_X$ and $f_Y$ are the spatial frequencies in the $x$ and $y$ directions respectively, that depend upon the pixel pitch of the recording CCD array. Eq. (1) reconstructs the complex amplitude distribution at the magnified image plane, from which the amplitude and phase of the magnified image can be extracted. In the present case, since the image plane is situated at the recording plane (Fig. 1a), meaning $d=0$, the setup basically represents image plane digital holographic microscope and the numerical reconstruction process to extract phase object information reduces to Fourier fringe analysis [61]. The numerical reconstruction process to extract object ($\Phi_O$) and reference ($\Phi_R$) phase information involves Fourier transforming the holograms and then filtering the frequency information corresponding to the object alone and then inverse Fourier transforming the resultant filtered spectrum. In this process, reference beam can be replaced by a scalar [21, 59-60]. The phase measurement capability of the microscope using the described phase reconstruction technique was validated by using glass microspheres immersed in microscope oil as object. Accurate thickness profiles of the microspheres were obtained, indicating that the Fourier fringe analysis provide accurate phase values [65]. Phase subtraction ($\Delta\Phi = \Phi_O - \Phi_R$) gives the phase information of the objects (RBCs) alone, by nullifying the phase due to system-related aberrations. The phase ($\Delta\Phi$) is then unwrapped using the Goldstein's branch cut method [66] to get the continuous phase distribution $\Delta\Phi_{Un}$, which is used to compute the optical path length or optical thickness (POL) from the relation $OPL = \left(\frac{\lambda}{2\pi}\right)\Delta\Phi_{Un}$. OPL gives the thickness /height profile of RBC from the relationship $h(x,y) = OPL/\Delta n$, where $\Delta n = n_{RBC} - n_{saline}$, the refractive index difference between RBC and the saline solution. The constant average refractive index ($n_{RBC}$) of a healthy RBC and the saline solution ($n_{saline}$) are 1.42 [67] and 1.334, respectively.[68] The thickness profiles (2D, 3D and cross-sectional) of RBCs, computed from the reconstructed continuous phase distribution obtained after phase subtraction and using the mentioned refractive index values, are shown in Fig. 9 in Sec. 3.1.

The 3D thickness profile of the RBC is used to calculate various geometrical parameters such as surface area, volume, sphericity index etc. The volume of the cells is computed using the relation [69]

$$V = dA \sum_{i=1}^{N} h_i \tag{2}$$

where $dA$ is the area of each pixel on the 3D thickness profile of RBC considering the lateral magnification of the system and $h_i$ is the thickness at each pixel obtained from the optical phase map using the constant average refractive index of the cell and the medium surrounding it (blood plasma).

Surface area (SA) is the addition of the projected area ($A_p$) and curved surface area of the cell. It can be written as [69]

$$SA = dA \sum_{i=1}^{x} \sum_{k=1}^{y} \sqrt{(1 + \delta h_x^2(i,k) + \delta h_y^2(i,k))} + A_p \tag{3}$$

$\delta h_x$ and $\delta h_y$ are the gradients of thickness along the x and y direction of the cell thickness profile. $i$ and $k$ correspond to the position of the pixel on 3D thickness profile of the cell. $A_p$ is given by $A_p = N \frac{\Delta x^2}{M^2}$, where $N$ is the number of pixels occupied by the cell in a plane, $\Delta x$ is the pixel pitch (in this case 4.65μm) and $M$ is the lateral magnification of the imaged object at the image (or hologram) plane (in the present case 24.17).

## 2.2 Parametric models for RBC shape

Our primary concern is to examine whether the RBC parametric models, which are most frequently used in numerical simulation studies and the diagnostic techniques (like cytometry), give an estimate of the geometrical shape and the key quantities (volume, SA and SI) of healthy RBCs that is in close agreement with the DHM results. Therefore, from the several existing parametric models[1,26-37], we have chosen the Cassini model[26,27], Fung and Tong (FT) model[28-30], Skalak model[31,32], Kuchel-Fackerell (KF) model[35,36], San Martin-Sebastian-Sancho-Alvarez (SMSSA) model[37] and, Yurkin model[1]. A detailed description of these models is given in the Supplementary Information. The diameter (*d*), maximum thickness (*t_max*), and minimum thickness (*t_min*) are the essential input parameters in these models. The values of each of these parameters for a single RBC are chosen in such a way that they lie within the experimentally reported range of their normal distributions. (Fig. S7 Supplementary Information) All the calculations were done in MATLAB. The volume, SA, and SI of the RBC shapes generated using the Cassini model are calculated using the formulae given by Angelov and Mladenov [28] (see Supplementary Information). For the other models, the volume and the SA are computed through the standard numerical integration procedure. SI is calculated using the definition [70], $SI = 4.84 \frac{V^{\frac{2}{3}}}{SA}$.

## 3. RESULTS AND DISCUSSION

### 3.1 RBC thickness profile and Phase map

Representative reference and sample holograms recorded using the DHM set up described in Sec. 2.1.2 are shown in Fig. 1. The reconstructed false-coloured phase images obtained using the method explained in Sec. 2.1.3 are shown Fig. 2.

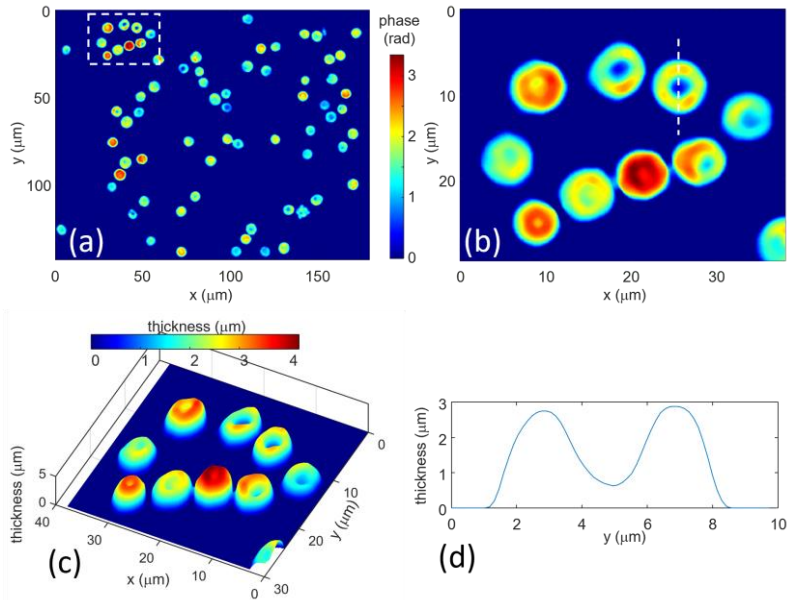

**FIGURE 2**: (a) False-coloured phase images of, (a) the object hologram in Fig. 1(c), (b) a selected segment of the object hologram in Fig. 1(d), (c) 3D thickness profile corresponding to the phase profile in (b), (d) cross-sectional thickness profile of a selected RBC

To assess how well the parametric models can describe the biconcave discocyte shape of healthy RBCs, we first compare the 2D meridional cross-section of a cell obtained using the DHM and the models. Fig. 3 shows representative results for the half meridional cross-sectional profiles with the values of diameter, minimum thickness ($t_{min}$) and maximum thickness ($t_{max}$) in the normal range for healthy RBCs. The applicability of a model relies on its ability to mimic the central discoid region, a characteristic region with minimum thickness, and the outer lobes defining the maximum thickness of the cell. It is evident from the Fig.3 that the thickness profiles given by the Skalak and the KF models, are closest to the DHM profile.

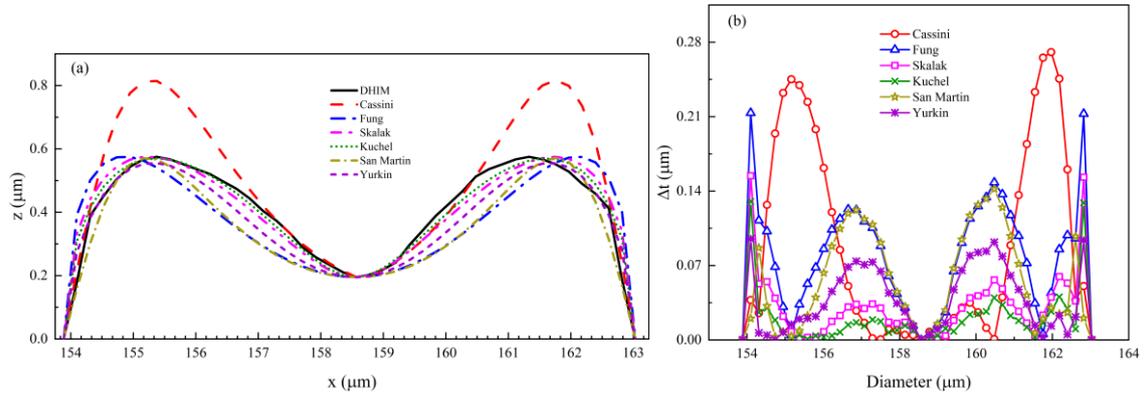

**FIGURE 3**: (a) Half meridional cross-section of a healthy RBC obtained using DHM and different models, (b) Deviation in the cell thickness with respect to DHM for different models.

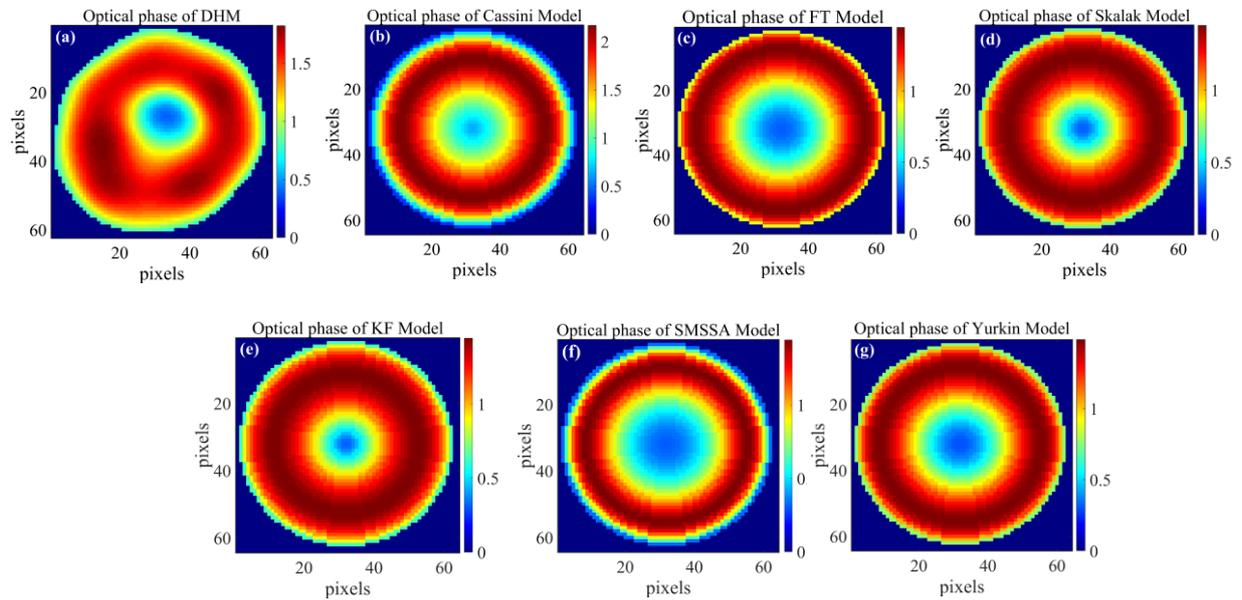

**FIGURE 4**: Optical phase maps for a healthy RBC, (a) DHM, (b) Cassini model (c) FT model and, (d) Skalak model, (e) KF Model, (f) SMSSA Model, (g)Yurkin Model. The axes scales are kept uniform in all the maps for correct comparison.

The absolute deviation in the meridional thickness profile of the cell with respect to the DHM profile (Fig. 3(b)) suggests that the KF model provides the best geometrical description of the healthy RBC shape. The large deviations in the thickness profile of Cassini model with only two parameters compared to the other multi-parametric models highlight the importance of the inclusion of as many characteristic geometrical parameters as possible in a model for the RBC shape. The comparison of the 2D meridional thickness profile generated by a model with the experimental profile gives only a first-hand estimate of its ability to generate an appropriate equilibrium RBC shape. A better way to judge a parametric model would be the comparison of the theoretically calculated optical phase map from the modeled 3D thickness profile to the actual experimental optical phase map. We use the relation $\Delta\Phi_{Un} = \left(\frac{2\pi}{\lambda}\right) \Delta n\, h(x,y)$ to obtain the optical phase map, where $h(x,y)$ is 3D thickness profile of the cell generated by a parametric model. $\Delta n$, the difference of the refractive indices of the medium inside the cell ($n_{RBC}$=1.42), and the external saline solution ($n_{saline}$=1.334), is considered to be uniform throughout the volume of the RBC. Basically, it is an inversion of the QPI problem of extraction of thickness profile from the optical phase recorded in the digital holograms. The representative phase maps for a healthy RBC, obtained using DHM and the models, are shown in Fig. 4. The DHM optical phase maps of more RBCs are given in the Supplementary information (Fig. S8). The characteristic parameters, $d$, $t_{max}$, and $t_{min}$ used in the model calculations lie within the range of experimentally reported values. Their values are chosen in a way that they give the closest estimate of the optical phase map in comparison to the experimental result. The color maps clearly show that the KF model (Fig. 4(e)) and the Skalak model (Fig. 4(d)) quite closely reproduce all the essential geometrical features of a healthy RBC i.e. the central disc region of minimum thickness, the lobes with maximum thickness and the intermediate curved region of gradually increasing thickness, as observed in the DHM phase map (Fig. 4(a)). While comparing the results of the models with the DHM results, one should keep in mind that the models give perfect parametric curves and surfaces without any local undulations that exist on the cell membranes whereas these undulations are captured in DHM. The ability of the KF model and the Skalak model to give a better overall account of the 3D RBC shape relative to other models points to the importance of the consideration of biomechanical properties of the cell membrane (mainly membrane elasticity and bending energy) in the parametrization of the shape model. The parametrization of the mathematical expression of the KF model is close to the a model proposed by Svetina and Zeka[71], based on the minimization of the bending energy of a bilayer membrane. The Skalak model considers various elastic moduli of the cell membrane as a single strain function.[32]

The ability of the parametric models to give the optical phase maps comparable to the DHM results is significant from the viewpoint of the RI-thickness coupling problem in the QPI methods using single-shot DHM. The coupling of the geometrical thickness of the cell and the refractive index of the cell medium makes it difficult to extract the two independently from each other.[50] One of the several approaches for decoupling the refractive index from the thickness is to extract the integral RI from the thickness profile obtained using an approximated model.[56] This approach involves fitting a reference shape (sphere or ellipsoid) to the phase map of a cell to obtain the RI profile. In this context, the use of the RBC shape models instead of the sphere or ellipsoid would give more realistic information about the cellular refractive index and the inhomogeneity of the cellular fluid. To ascertain this, we generate the refractive index map using the relation, $n_{RBC}(x,y) = n_{saline} + \left(\frac{\lambda}{2\pi}\right)\frac{\Delta\Phi_{Un}(x,y)}{h(x,y)}$, where $\Delta\Phi_{Un}(x,y)$ is the unwrapped optical phase map obtained from DHM, and $h(x,y)$ is the RBC thickness profile from a parametric model. The representative results for the Skalak model and the KF model are shown in Fig. 5 along with the common-path diffraction optical tomography (cDOT) results [53]. It is evident that the integral RI profile extracted from the holographic optical phase using the model thickness profile closely resembles the RI profile from cDOT experiments. Similar RI profiles have also been reported in high-resolution 3D imaging of healthy RBCs using optical diffraction tomography.[51,53,54] Thus, the RBC parametric models offer a simpler solution to the RI-thickness decoupling problem and could be very useful for the biochemical characterization of the inhomogeneities in the cellular fluid.

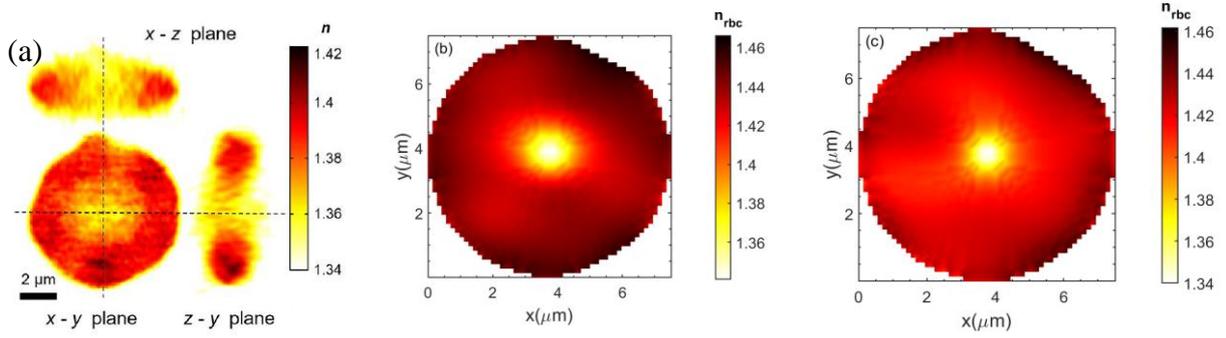

**FIGURE 5 :** (a) cross-sectional images (along the x-y, the z-y and the x-z planes) of the 3-D RI tomogram of a RBC [Ref. 53]; Integral RI profile of a RBC extracted from the DHM optical phase map and the thickness profile generated using (b) Skalak model, and (c) KF model;

## 3.2 RBC geometrical parameters, its distributions and variability

For 500 healthy RBCs investigated using DHM with a diameter in the range of 6.0-9.0 µm, the average diameter has been found to be 7.43 µm. The average maximum thickness ($t_{max}$) is found to be 3.23 ± 0.48 µm which is closer to a reported value, 3.12 ± 0.47, in a scanning flow cytometry study[2]. Considering the RBC surface undulations, the average of the mean thickness across the surface of all the cell is 2.0 ± 0.26 µm. RBC volume, SA area and SI, which are important for the classification of RBC subpopulations and diagnosis, are calculated according to the methods explained in Sec. 2.1 and 2.2. In the case of models, the essential input cell parameters i.e. diameter, $t_{min}$, $t_{max}$, eccentricity (in Cassini model) for 500 cells, have been chosen in such a way that their values lie within experimentally reported ranges of their normal distributions. The variations in the RBC volume, SA and SI with the diameter are shown in Fig. 6(a), 6(b), and 6(c), respectively. It can be observed that the results of DHM and the models show qualitatively similar variations in volume and SA with diameter (increase in V and SA with increase in *d*) whereas the variation in SI differs significantly for the different models. It can be gauged from Fig. 6(a) and 6(b) that the variations in *V* and *SA* with the diameter are linear for DHM results and FT, KF, SMSSA models whereas it follows power laws in case of Cassini, Skalak and Yurkin models. The linear correlations of *V* and *A* with the diameter indicate that the cell shapes cannot vary independent of the volume.[70] Its implications in the results of SI can be seen in Fig. 6(c). While SI shows variation around a constant mean value for the Cassini, Skalak and Yurkin models, it exhibits negative correlation with the for the FT, KF and SMSSA models as observed in the DHM results too.

The negative correlation between the SI and diameter observed in DHM is in agreement with the results reported by Canham and Burton.[70] It has also been argued that if a constant SI is the link between *SA* and *V*, then *SA* would be proportional to $V^{2/3}$, and the *A* vs. *V* curve would be concave downward. To ascertain this, we have obtained the *SA-V* curves as shown in Fig. 7. We find that $SA \propto V^{0.6}$ for the Cassini, Skalak and Yurkin models which show a nearly constant mean SI with diameter. The FT, KF and SMSSA models as well as the DHM results show positive linear correlation between *SA* and *V*. The constant mean SI with increasing diameter implies that the shape of RBCs does not vary significantly whereas the negative correlation between SI and diameter indicates decreasing sphericity of the cells due to changes in the characteristic regions of the RBC shapes i.e. the central discoid region and the peripheral lobes. Considering the variability of the shape and size of the normal RBCs within the diameter range of 6.0 – 9.0 μm, SI – diameter negative correlation is more plausible compared to the constant mean SI scenario. As SI is considered to be more sensitive marker for the discrimination among different RBC types [2], the choice of parametric models for the numerical simulations could have a significant bearing on the results where the RBC shape plays an important role.

To get more quantitative insight of the RBC geometrical parameters acquired through the models and its validity vis-à-vis the DHM results, we perform descriptive statistical analysis using the distribution plots and the box plots of the normalized counts of the surface area,

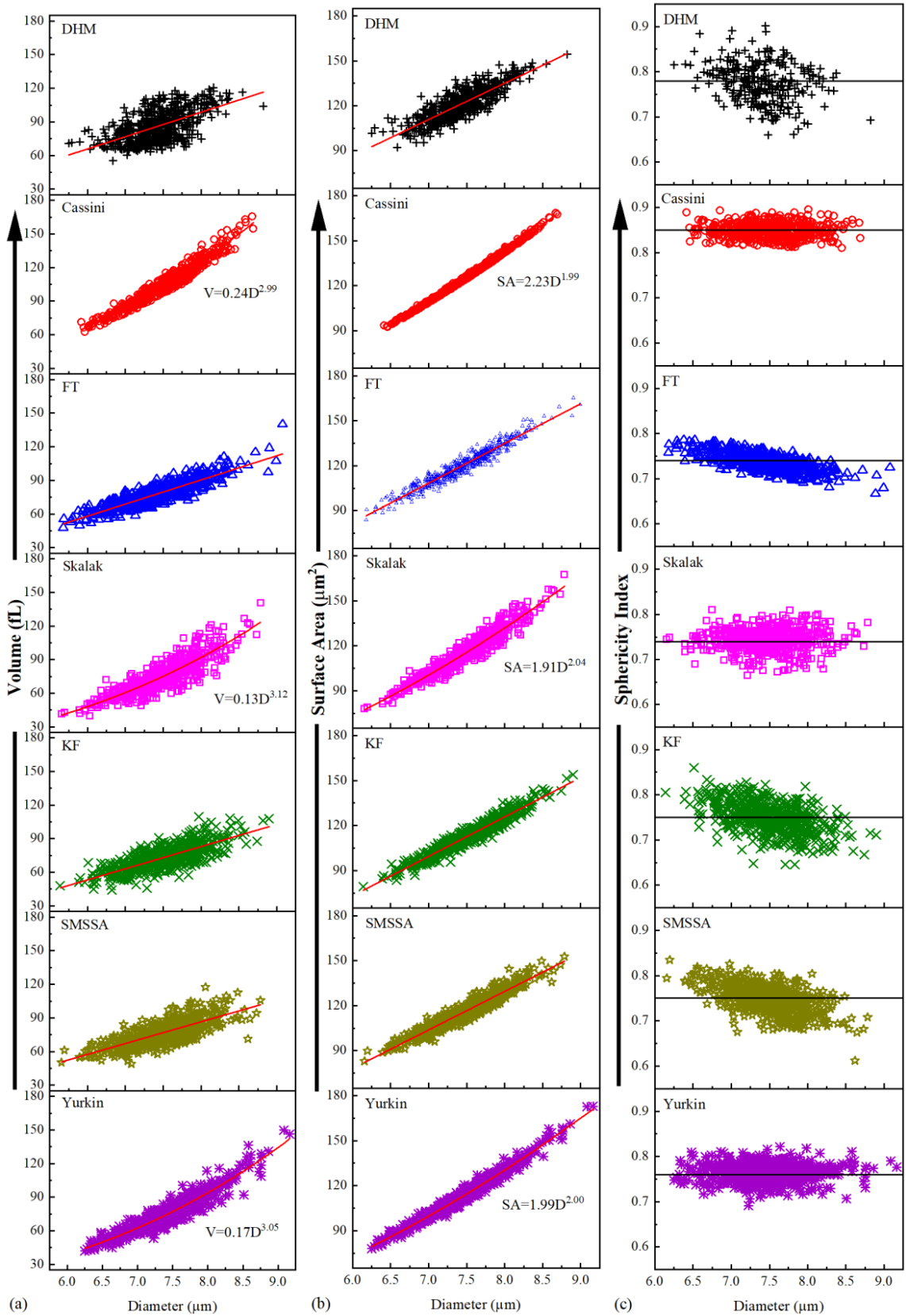

**FIGURE 6:** (a) Volume, (b) Surface Area, and (c) Sphericity Index of 500 healthy RBCs from DHM and different parametric models. The lines correspond to the linear fit and the power-law fit in the respective cases in (b). The equations for power-law fit are displayed in the graphs wherever applicable. The horizontal lines in the graphs in (c) correspond to the mean value of SI.

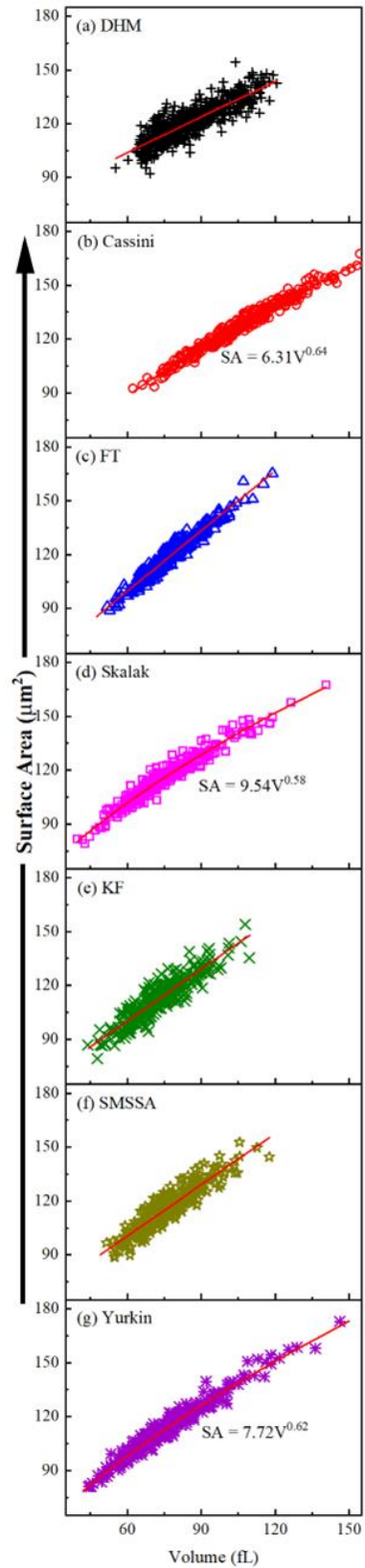

**Figure 7:** Correlation between surface area and volume for (a) DHM, (b) Cassini model, (c) FT model, (d) Skalak model, (e) KF model, (f) SMSSA model and, (g) Yurkin model. The lines correspond to the linear fit and the power-law fit in the respective cases. The equations for power-law fit are displayed in the graphs wherever applicable.

the volume and the sphericity index of 500 healthy RBCs. Fig. 8 and Fig. 9 give the distribution plots and the box plots, respectively. All the distributions are normal (Gaussian) and it should be noted that the box range in the descriptive statistics corresponds to one standard deviation ($\sigma$) about the mean value. The whiskers, which extend up to 1.5$\sigma$, give the extent of the spread of the maximum number of data points that indicates the variability of the given parameter. The mean values of the surface area, the volume, and the sphericity index along with the standard deviation are listed in Table 1. These values are within the range of experimentally reported results for healthy RBCs. Keeping the distribution plots for the Cassini model aside for a moment, some general observations from the distribution curves and the box plots for the DHM and the other five models are: (i) the mean values of the volume, SA, and SI for the models are a little lower than the DHM results, (ii) while the volume distribution curves significantly overlap with nearly the same mean value, the SA distribution curves are more distinct, especially in the region with SA lower than the mean value, (iii) the SI distributions exhibit the most distinct curves with different mean values and the distribution widths. These observations elucidate the impact of different parametrization of RBC shape in different models on the variability of the volume, SA and SI. It can be easily corroborated through the results of the Cassini model. To understand it, we should note that the diameter (*d*) and the eccentricity ($\varepsilon_p = 1 - \varepsilon$) are two input parameters in the Cassini model. As explained in Sec. 1.1 of Supplementary

**TABLE 1**: Geometrical parameters of the healthy RBCs determined using Experiment and analytical models.

|  | Volume [fL] | Surface Area [$\mu m^2$] | Sphericity Index |
|---|---|---|---|
| **DHM** | 86.1 ± 14.0 | 121.2 ± 10.8 | 0.78 ± 0.04 |
| **SFC (Ref. 2)** | 82.2 ± 21.6 | 104.3 ± 20.4 | 0.82 ± 0.08 |
| **cDOT (Ref. 16)** | 94.8 ± 11.4 | 137.8 ± 13.3 | 0.73 ± 0.05 |
| **cDOT (Ref. 53)** | 90.5 ± 11.4 | 144.1 ± 17.4 | 0.68 ± 0.06 |
| **Cassini** | 104.8 ± 18.6 | 126.2 ± 14.7 | 0.85 ± 0.02 |
| **FT** | 78.68 ± 11.49 | 120.50 ± 13.13 | 0.74 ± 0.02 |
| **Skalak** | 75.39 ± 15.44 | 115.47 ± 13.98 | 0.74 ± 0.03 |
| **KF** | 73.18 ± 11.26 | 112.83 ± 11.95 | 0.75 ± 0.03 |
| **SMSSA** | 76.46 ± 11.09 | 116.14 ± 11.57 | 0.75 ± 0.03 |
| **Yurkin** | 76.55 ± 16.31 | 113.42 ± 15.08 | 0.76 ± 0.02 |

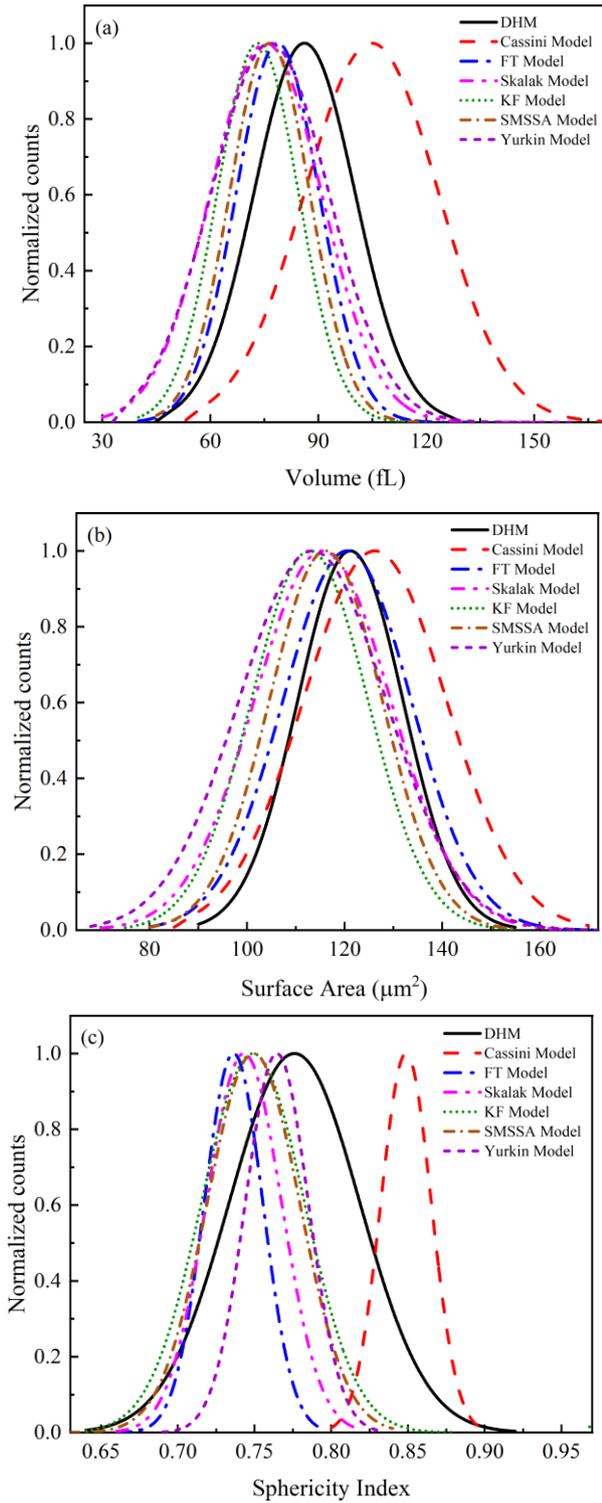
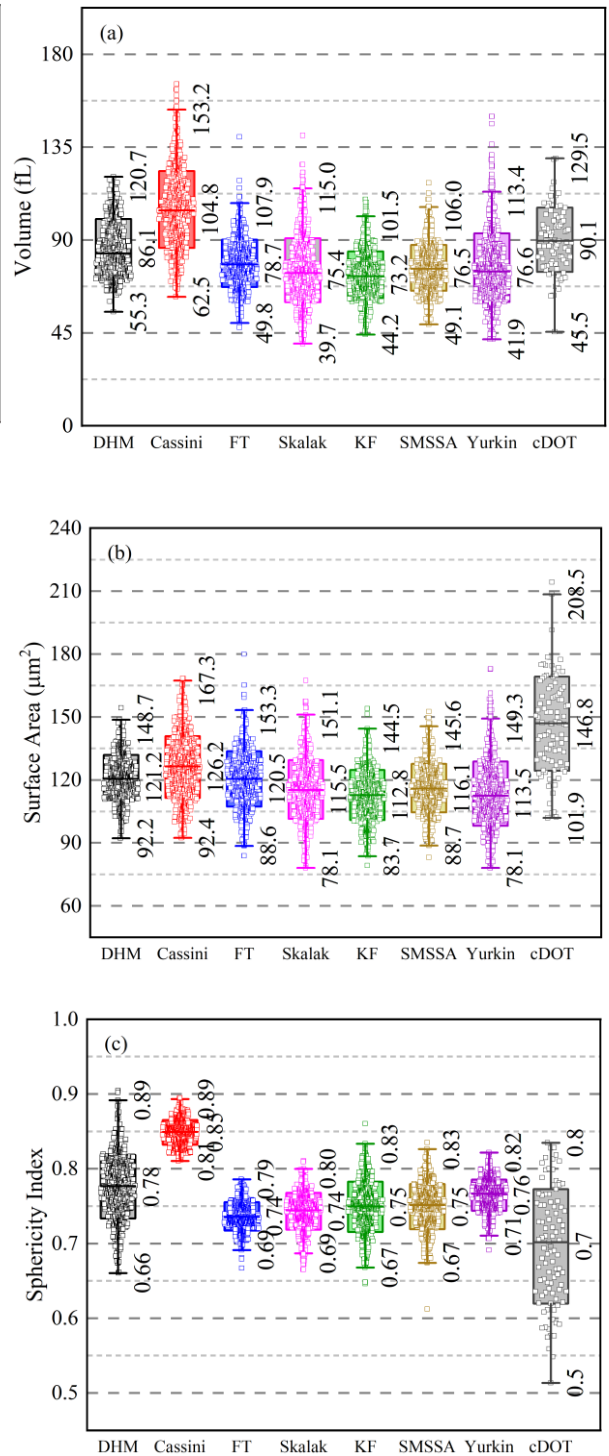

**FIGURE 8:** Distributions of the normalized counts of (a) surface area, (b) volume and, (c) sphericity index of 500 healthy RBCs

**FIGURE 9:** Box plots for (a) surface area, (b) volume and, (c) sphericity index of 500 healthy RBCs. The values in the middle of the box are the mean values.

Information, $\varepsilon_p$ relating two geometrical parameters *a* and *c*, governs all the essential characteristic parameters such as *d*, $t_{min}$, $t_{max}$ defining the shape and size of a normal RBC. Fig. 10 demonstrates that the the range of $\varepsilon_p$ value is severely restricted below 0.90 for generating an RBC shape with the values of $d$, $t_{min}$, $t_{max}$ in the normal range. It also points towards the limitations of the Cassini model in generating the central discoid region compared to other models and the same is further validated by our results of the thickness profile and the optical phase profile of normal RBC shown in Fig. 3 and 4 respectively. Our results in Fig. 8(c) and Fig. 9(c), showing the range of SI to be 0.8 – 0.9, suggest a close correspondence between SI and $\varepsilon_p$. The effect of the difference in parametrization in RBC models on the variability of the volume, SA and SI can also be seen in the results of the FT and Yurkin models. Our results of the variation in the volume (Fig.6(a)), SA (Fig.6(b)) and SI (Fig.6(c)) with the cell diameter suggest that the inclusion of a parameter $\varepsilon = T_{max}/d$ in the Fung model [28-30] by Yurkin et al [1] (See Sec. 2.2.6, Supplementary Information) leads to a qualitative and quantitative change in the dependence of the volume, SA and SI on the diameter of the cell. The effect of the parametric difference between the FT and Yurkin models is more clearly visible in the distribution and box plots for the SI in Fig. 8(c) and 9(c), respectively.

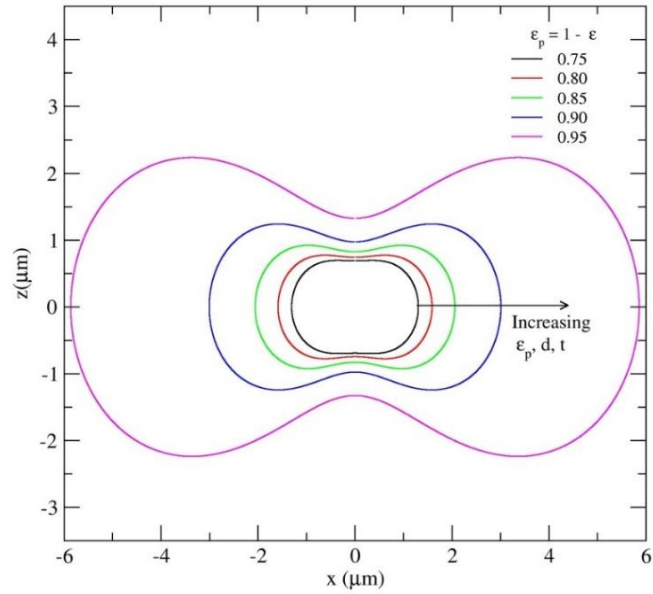

**FIGURE 10:** Cassinian curves for different $\varepsilon_p$ values. The curves are derived using Eq. (1) described in the Supplementary Information. The right arrow indicates the increase in the cell diameter and thickness with the increase in the $\varepsilon_p$.

## 4. CONCLUSIONS

Present investigations of the geometrical shape and size of healthy human RBCs using DHM experiments and six most frequently used parametric RBC shape models lead us to the following conclusions:

(i) In addition to the meridional thickness profiles, a comparison of the optical phase maps, generated from the 3D thickness profiles given by the models, with the DHM results

(i) enables a better assessment of the appropriateness of the models for the realistic mimicking of the biconcave discocyte shape of healthy RBCs.

(ii) the RBC parametric models offer a simpler solution to RI-thickness decoupling problem in QPI methods. Coupling the thickness profile from the parametric models with the DHM optical phase provides the cell refractive index map, which can give many features like structure, structure coherence, and its time variation. It will enhance the RBC classification probability of DHMs.

(iii) The observed greater variability of SA and SI compared to the volume in DHM results highlights the variability of the RBC shapes in healthy human blood.

(iv) The six investigated RBC shape models can be classified into two categories based on the nature of the variation in V and SA with the cell diameter. Cassini, Skalak, and Yurkin models exhibit the power-law dependence of V and SA on the diameter. FT, KF, and SMSSA models show a linear variation of V and SA with the diameter, which is also in agreement with the DHM result.

(v) While the Cassini, Skalak, and Yurkin models yield RBC shapes with nearly constant mean SI independent of the cell diameter, the negative SI – diameter correlation observed in DHM results and the FT, KF, SMSSA models suggests that the latter models would be more appropriate and correct choice in the numerical simulations where SI is an important marker for the discrimination between the RBCs of different shapes.

(vi) The test of the appropriateness of a parametric RBC shape model for its application in a numerical simulation warrants an assessment of the thickness profile as well as its success in giving a reasonable account of the variability of the volume, SA and SI through statistical distributions.

**ACKNOWLEDGMENTS**

Authors acknowledge the use of computational facilities at the Department of Physics, Sardar Patel University (SPU) established through DST-FIST program of Department of Science and Technology, Govt. of India and the PARAM SHAVAK supercomputing facility established by Govt. of Gujarat at SPU. AA acknowledges the research grant from SERB (EMR/2017/002724). Authors are grateful to Dr. Niteshkumar S. Mistry for providing the guidance in sample preparation methods using the facilities at the Department of Microbiology, Arts, Science & R.A. Patel Commerce College, Bhadran, Gujarat.